\documentclass[journal]{IEEEtran}
\usepackage{cite}
\usepackage{xcolor}
\usepackage{soul}
\usepackage[utf8]{inputenc}

\ifCLASSINFOpdf
   \usepackage[pdftex]{graphicx}
\else
\fi

\usepackage{amsmath}
\usepackage{upgreek}
\usepackage{array}

\ifCLASSOPTIONcompsoc
  \usepackage[caption=false,font=normalsize,labelfont=sf,textfont=sf]{subfig}
\else
  \usepackage[caption=false,font=footnotesize]{subfig}
\fi

\ifCLASSOPTIONcaptionsoff
  \usepackage[nomarkers]{endfloat}
 \let\MYoriglatexcaption\caption
 \renewcommand{\caption}[2][\relax]{\MYoriglatexcaption[#2]{#2}}
\fi
\usepackage{url}

\newcommand\circone{$\raisebox{.5pt}{\textcircled{\raisebox{-.9pt} {1}}}$ }
\newcommand\circtwo{$\raisebox{.5pt}{\textcircled{\raisebox{-.9pt} {2}}}$ }
\newcommand\circtwons{$\raisebox{.5pt}{\textcircled{\raisebox{-.9pt} {2}}}$}
\newcommand\circthree{$\raisebox{.5pt}{\textcircled{\raisebox{-.9pt} {3}}}$ }

\begin{document}
%
\title{Residual Phase Noise Measurement of Optical Second Harmonic Generation in PPLN Waveguides
}
%
%

\author{Marion Delehaye, Jacques Millo, Pierre-Yves Bourgeois, Lucas Groult, Rodolphe Boudot,  Enrico Rubiola, Emmanuel Bigler, Yann Kersalé and Clément Lacroûte
\thanks{
	All the authors are with FEMTO-ST institute, univ. Bourgogne Franche-Comté, CNRS, ENSMM Time and frequency dept. 26 Rue de l'\'Epitaphe, 25030 Besançon cedex, France.}
}

%
%

\markboth{}%
{Shell \MakeLowercase{\textit{et al.}}}
%



\maketitle

\begin{abstract}
We report on the characterization, including residual phase noise and fractional frequency instability, of fiber-coupled PPLN non-linear crystals. These components are devoted to frequency doubling 871~nm light from an extended-cavity diode laser to produce a 435.5~nm beam, corresponding to the ytterbium ion electric quadrupole clock transition. We measure doubling efficiencies of up  to 117.5\%/W. Using a Mach-Zehnder interferometer and an original noise rejection technique, the residual phase noise of the doublers is estimated to be lower than ${\bf -35\, dBrad^2/Hz}$ at 1 Hz, making these modules compatible with up-to-date optical clocks and ultra-stable cavities. The influence of external parameters such as pump laser frequency and intensity is investigated, showing that they do not limit the stability of the frequency-doubled signal. Our results demonstrate that such compact, fiber-coupled modules are suitable for use in ultra-low phase noise metrological experiments, including transportable optical atomic clocks.
\end{abstract}


%
\IEEEpeerreviewmaketitle

\section{Introduction}
%
%
%
%
\IEEEPARstart{O}{ptical} atomic clocks have reached unprecedented fractional frequency instabilities and accuracies, as low as $6\times 10^{-17}\tau^{-1/2}$ for the former~\cite{Schioppo_2016} and $3\times 10^{-18}$ for the latter~\cite{Bloom_2014}. In recent years, tremendous progress has been achieved in this domain thanks to improved performances of the clock lasers, which are pre-stabilized using ultra-stable Fabry-Perot resonators~\cite{Jiang_2011}. Many state-of-the-art optical clocks rely on a cavity-stabilized laser, frequency-doubled via second harmonic generation (SHG) to produce the laser clock frequency~\cite{Chou2010,Tyumenev_2016}. Other applications requiring SHG and low phase noise at optical frequencies include precision spectroscopy, optical frequency combs \cite{Tyumenev_2016}, and interferometric gravitational waves detectors \cite{Yeaton_Massey_2012}. For all of those applications, it is necessary to ensure that the SHG setup does not limit the optical signal frequency stability.

In this context, bulk crystals and periodically poled (PP) potassium titanyl phosphate (KTP) or potassium niobate (KN) crystals have been used and characterized. It has been shown that they introduce very little additional phase noise, but exhibited relatively poor intrinsic SHG efficiency~\cite{Wynands1995,Stenger2002,Yeaton_Massey_2012}. More recently, fiber-coupled periodically poled Lithium Niobate (PPLN) waveguides have found applications in ultra-cold atom interferometers~\cite{Leveque2014}. These devices have shown to exhibit very good conversion efficiencies in a compact and robust package, suitable for transportable or even space applications.

Several transportable optical atomic clocks are currently being developed towards the future implementation of large optical clock networks. These ensembles will impact fundamental physics tests, relativistic geodesy, and time and frequency metrology~\cite{Riehle2017}. Both transportable single-ion and optical-lattice clocks are being developed ~\cite{Cao2017, Koller2017, Lacroute2016}, many of which would benefit from the use of compact and efficient SHG modules.

In this article, we characterize such SHG modules based on fiber-coupled PPLN waveguides in terms of residual phase noise and fractional frequency stability. These modules are compact, highly-efficient, and avoid the use of an enhancement cavity. The frequency doubling stage is performed from a commercially-available extended cavity diode laser (ECDL) at a wavelength $\lambda = 871\,{\rm nm}$ to generate a beam at $\lambda = 435.5\,{\rm nm}$, corresponding to the ${\rm Yb^+}$ electric quadrupole clock transition.

\begin{figure}[]
	\centering
	\begin{minipage}{0.48\linewidth}
		\centering
	\includegraphics[trim=0 0 0 42, clip, width=.95\linewidth]{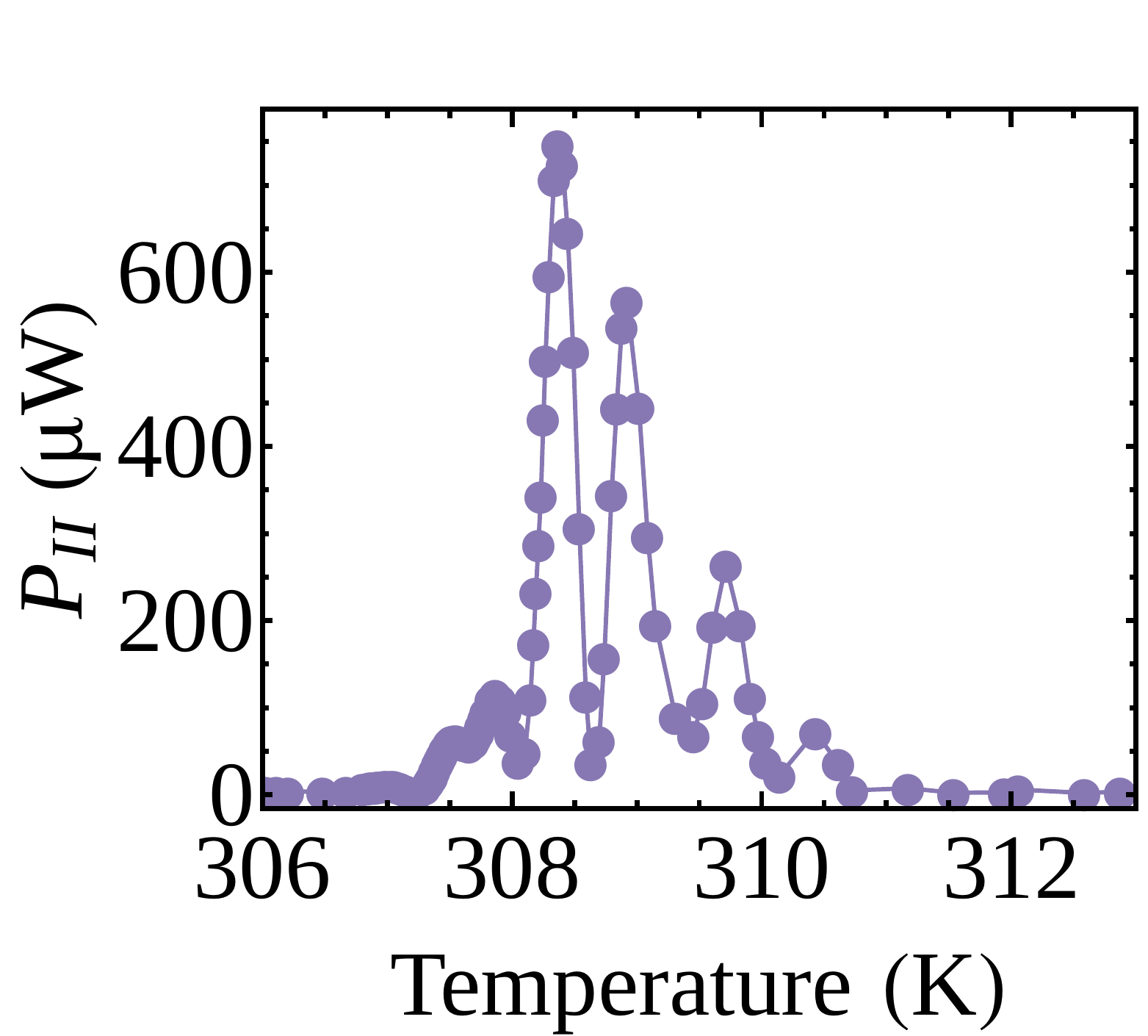}\\
	(a) 
	\end{minipage}
	\hfill
	\begin{minipage}{0.48\linewidth}
		\centering
	\includegraphics[trim=0 -10 0 42, clip, width=.95\linewidth]{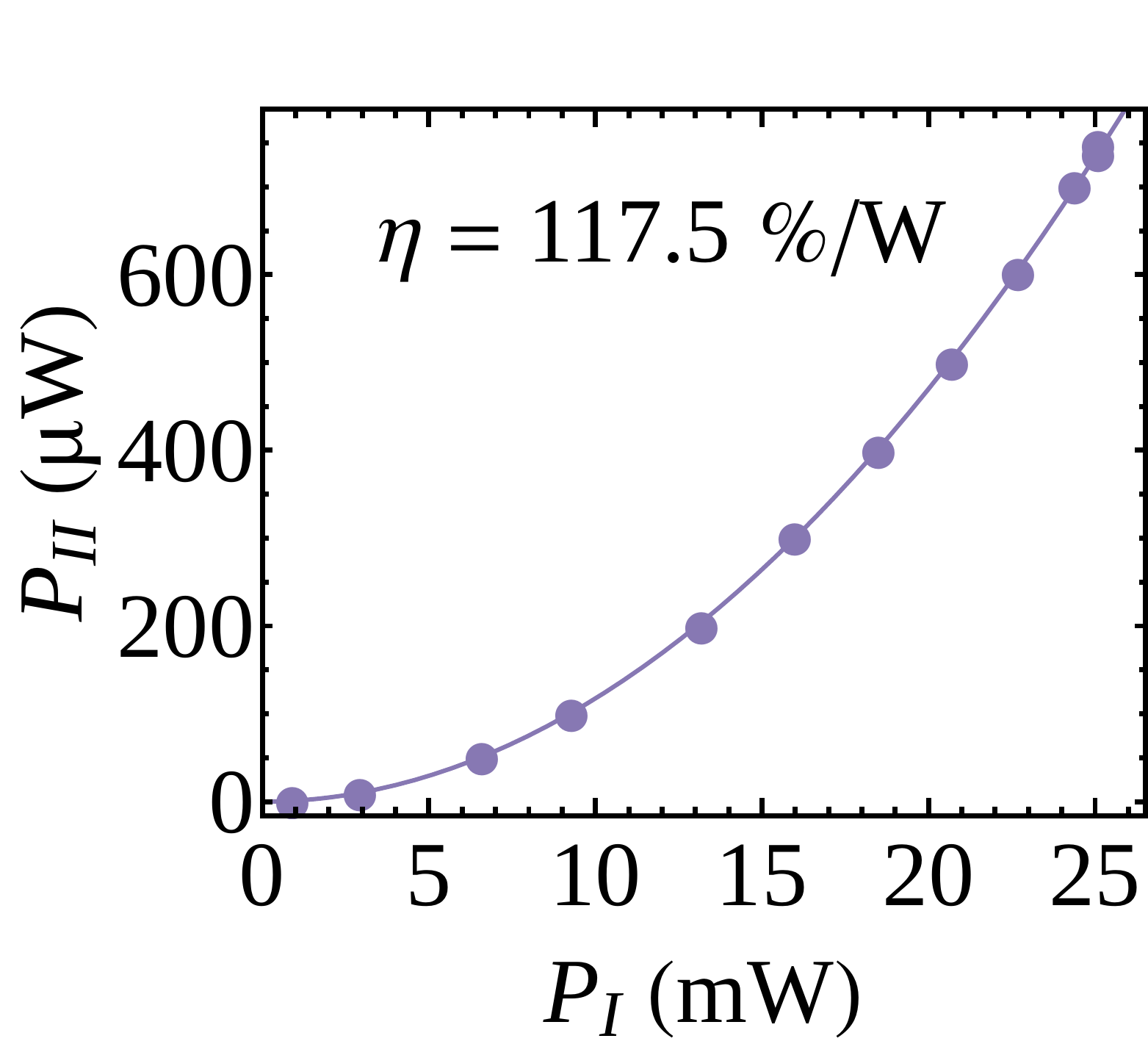}\\
	(b)
	\end{minipage}
	\caption{(a) Output power $P_{I\!I}$ as a function of the crystal temperature in K. The line is only here to guide the eyes. (b) Output power $P_{I\!I}$ as a function of the pump power $P_I$. Circles: datapoints. Line: parabolic fit to the data.}
	\label{fig:properties}
\end{figure}

The SHG modules we use are developed by NTT Electronics. They are composed of a PPLN-waveguide with a core section of about $8\times 8\,{\rm\upmu m^2}$ housed in a butterfly package with fiber-coupled input and electrical connections for thermistor and Peltier elements for temperature regulation. We have measured a conversion efficiency $\eta=P_{I\!I}/P_I^2$ of $117.5\,\%/{\rm W}$ for the best module, where $P_I$ is the pump laser power at 871~nm and $P_{I\!I}$ the output power signal at 435.5~nm at a working temperature of about 35~$^\circ$C (see Fig. \ref{fig:properties}).

\section{Phase-noise and instability measurements}
\label{sec:phasenoise}

\begin{figure}[b!]
	\centering
	\includegraphics[width=0.8\linewidth]{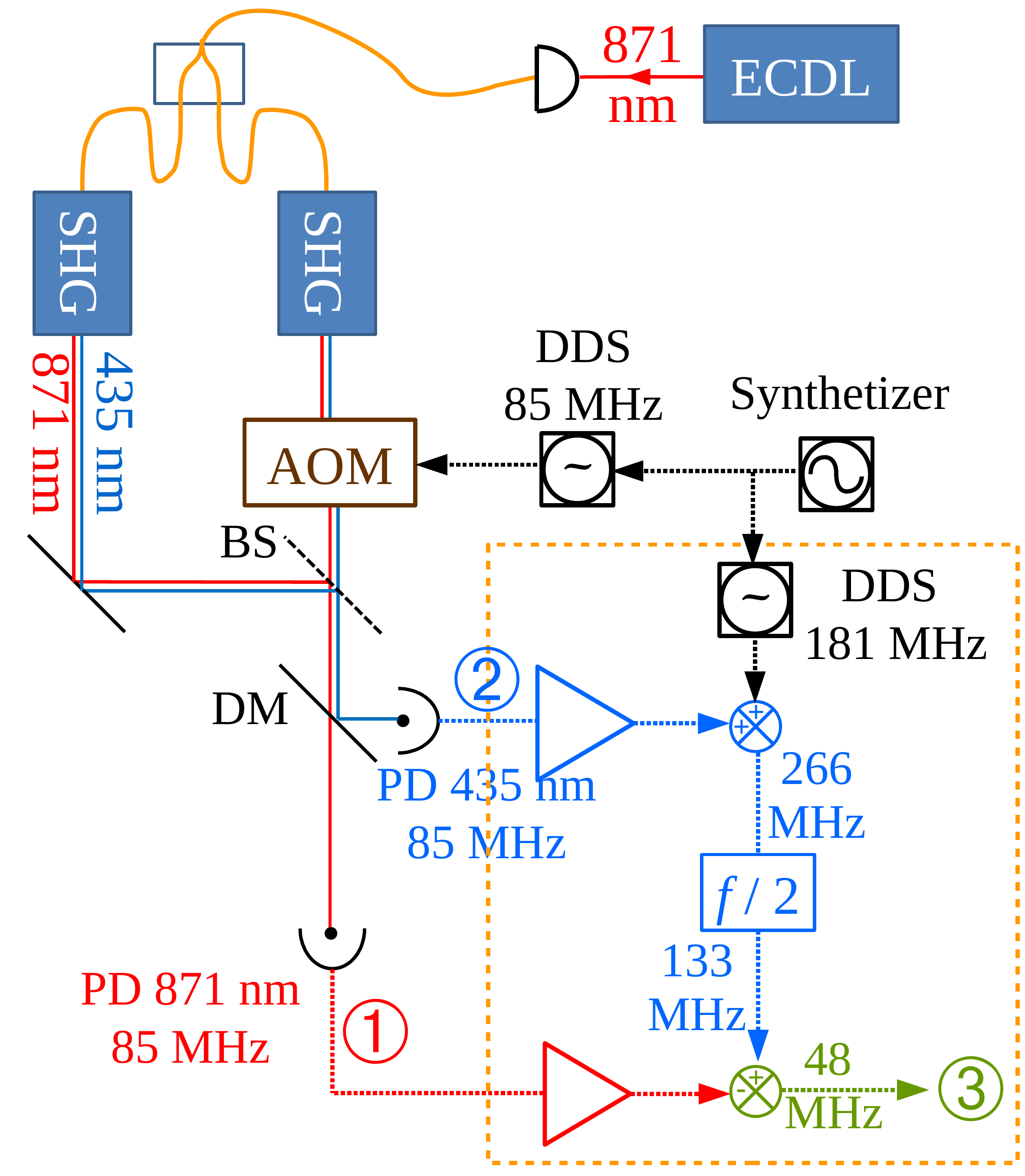}
	\caption{Optical and radio-frequency setup for phase noise and instability measurements. ECDL: extended cavity diode laser. Fiber length $\approx$ free-space length $\approx$ 1~m. SHG: second harmonic generation. AOM: acousto-optical modulator. DDS: direct digital synthesis. PD: photo-diode. DM: dichroic mirror}
	\label{fig:setup}
\end{figure}

The phase-noise measurement bench used to characterize the modules is described Fig. \ref{fig:setup}. It consists of a Mach-Zehnder interferometer; in one arm, the laser frequency is shifted using an acousto-optic modulator (AOM) driven at 85~MHz by a Direct Digital Synthesizer (DDS) clocked by a synthesizer referenced to an active hydrogen maser. The Mach-Zehnder interferometer is implemented by splitting the pump beam using a fiber splitter, and recombining both module outputs on a non-polarizing beam splitter. The interference between both arms results in a 85~MHz beatnote signal that carries any phase difference between the two paths, both for the infra-red (IR) and the blue signals at 871~nm and 435~nm respectively. The two wavelengths are separated using a dichroic mirror, and the IR and blue 85~MHz beatnote signals are sent to two different fast photodiodes (PD) and independently analyzed.

\begin{figure}[t!]
	\centering
	\includegraphics[width=1.\linewidth]{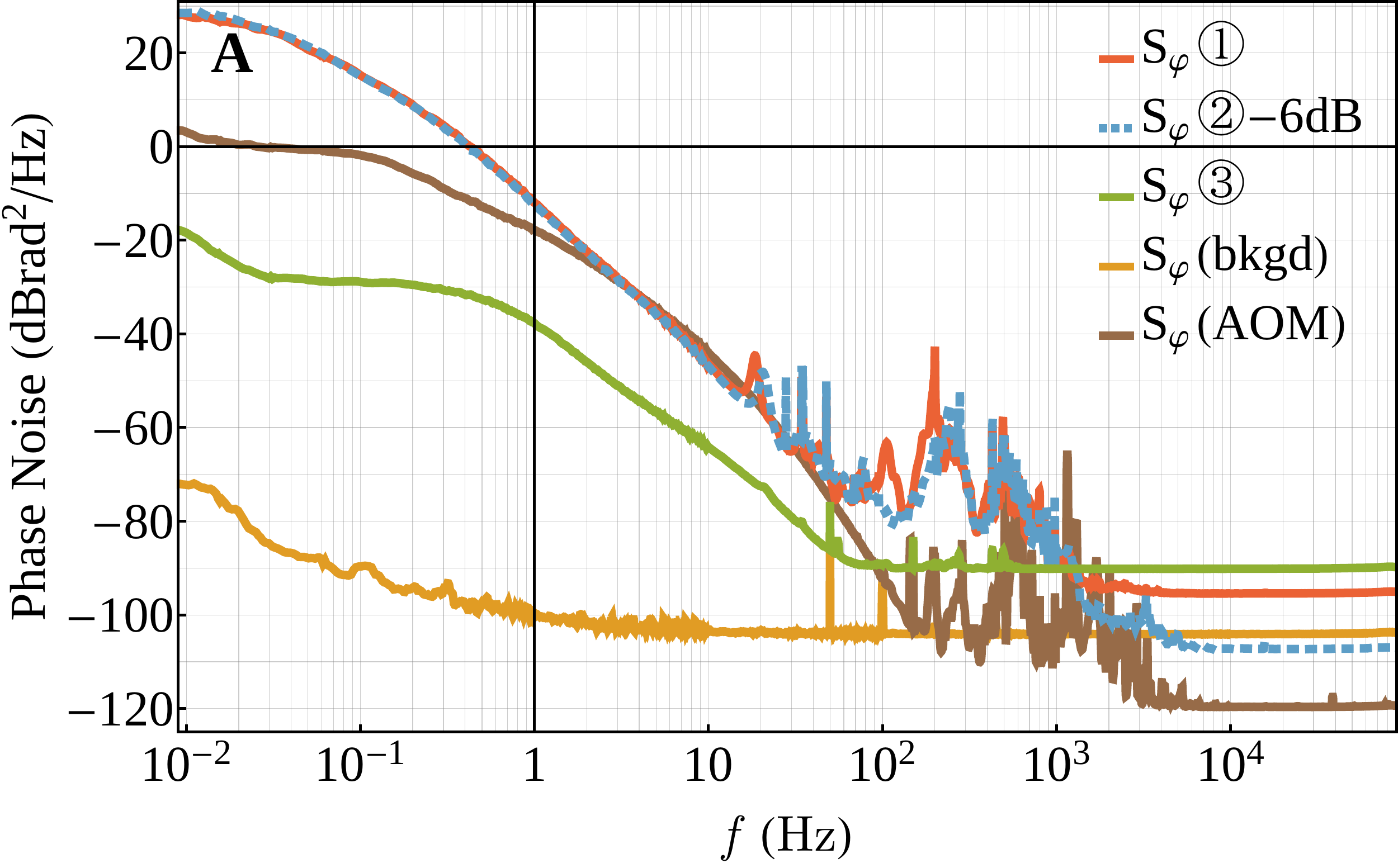}\\
	\includegraphics[width=.98\linewidth]{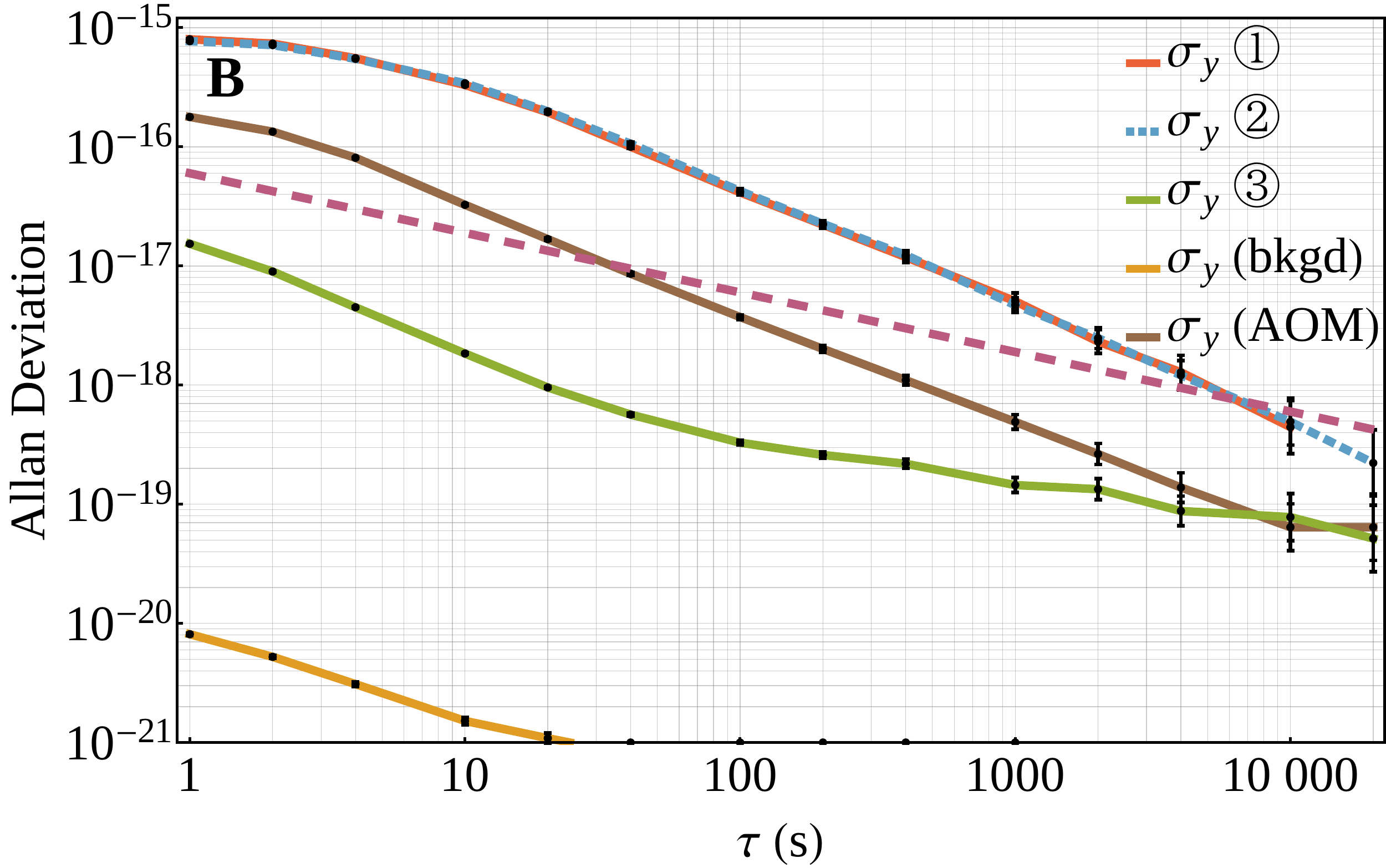}
	\caption{Phase noise spectral density ${\rm S}_\varphi(f)$ (A, top) and fractional frequency instability $\sigma_y(\tau)$ (B, bottom) from the same dataset of the RF signals. Red curve: 85~MHz signal \protect\circone obtained on the photodiode that collects 871~nm (344 THz) light. Dashed blue curve: 85~MHz signal \protect\circtwo obtained on the photodiode that collects 435.5~nm (688 THz) light, with ${\rm S}_\varphi(f)$ scaled to 871~nm. Green curve: 48~MHz signal \protect\circthree obtained after the common-mode noise-rejection bench (boxed in orange in Fig. \ref{fig:setup}). Orange curve: background (bkgd), obtained in \protect\circthree  when the output signals of the photodiodes are replaced by identical signals at 85~MHz generated by a DDS referenced to a hydrogen maser. Brown curve: AOM contribution measured at 871 nm. Purpled dashed curve, Fig. B: best published optical clock fractional frequency instability to date~\cite{Schioppo_2016}.} 
	\label{fig:phasenoise}
\end{figure}

In a first step, we have independently measured the phase noise power spectral density (PSD) ${\rm S}_\varphi (f)$ (Fig. \ref{fig:phasenoise}A)  and fractional frequency instability measured in terms of Allan deviation $\sigma_y (\tau)$ (Fig. \ref{fig:phasenoise}B) of the optical beatnotes at 871~nm and 435.5~nm (labeled respectively \circone and \circtwons). In this case, the photodiode output signal is amplified, low-pass filtered and measured using a phase-noise analyzer\footnote{Symmetricom 5125A} referenced to an active hydrogen maser. The measured phase-noise thus results from any optical length fluctuation in the beam path and from the putative noise induced by the PPLN crystal.

We have setup a second RF measurement bench, boxed by the orange-dashed line in Fig. \ref{fig:setup}. This setup rejects the phase noise coming from the interferometer optical length fluctuation and  sets an upper limit to the phase noise specifically generated by the frequency doubling process itself.
Here, the RF beatnotes from the 871~nm and 435.5~nm carriers are mixed in order to suppress any common noise between the two signals. For this purpose, the beatnote from the 435.5~nm signal is up-shifted to 266~MHz using a DDS. The 266~MHz signal is frequency divided by two to produce a 133~MHz signal, mixed at the end with the 85~MHz beatnote carried at 871~nm. This yields an output 48~MHz RF signal that can be analyzed with the phase-noise analyzer (green curve in Fig \ref{fig:phasenoise}A) and from which Allan deviation can be extracted (green curve in Fig \ref{fig:phasenoise}B). Any common noise on the 435.5~nm and 871~nm optical paths will be subtracted by the mixing process, so that this 48~MHz signal only carries information on the uncorrelated noise.
This includes noise generated by the SHG process itself, but also any optical length change that is not common-mode to the two wavelengths, potentially caused by differences in index of refraction in the SHG and AOM crystals, differences in thermal expansion of these crystals at the two wavelengths, or geometrical path differences. Nevertheless, the phase noise measured with this setup gives a more stringent upper bound to the phase noise induced by the SHG and allows to quantify the influence of the optical path length.

We have also measured the background phase-noise and fractional frequency instability 
by replacing the 85~MHz output signals of the photodiodes by identical signals generated by a DDS (orange curves in Fig \ref{fig:phasenoise}A and B). This shows that the phase-noise of the noise-rejection bench can be neglected with respect to the other sources of noise in the system.


For frequencies above 10 kHz, all phase noise measurements are found to be limited by the white noise floor of the RF amplifiers placed at the output of the fast photodiodes, given by $F k_{\rm B} T/P_{\rm in}$, where $F\approx 1$ dB is the noise figure, $k_{\rm B}T$ the thermal energy and $P_{\rm in}\approx -80\,{\rm dBm}$ the amplifier input power~\cite{Boudot_2012}. The various noise floors of Fig. \ref{fig:phasenoise}A correspond to different optical powers incident to the photodiodes.

We take away 6 dB from the phase noise PSD of \circtwo to take into account the factor two between the carrier frequencies of \circone and \circtwo; the phase noise spectra measured at 871~nm and 435.5~nm then almost perfectly match (red and dashed blue lines in Fig. \ref{fig:phasenoise}A), indicating mainly correlated noise, with a value  ${\rm S}_\varphi(1\,{\rm Hz})$ lower than $ -10\,{\rm dBrad^2 /Hz}$.
 
In Fig. \ref{fig:phasenoise}A, one can clearly identify acoustic noise peaks between 10 Hz and 5 kHz, the contribution of which could be lowered in a straightforward manner with acoustic shielding.
The spectrum slope is close to $f^{-4}$ for frequencies between 5~Hz and 70~Hz, and decreases below 5~Hz. $f^{-4}$ noise could stem from technical noise, including thermal fluctuations. We quantify the influence of other external parameters in a next part. The AOM contribution seems to be non-negligible, as shown below.

The  AOM  contribution to the measured phase noise  was  estimated  using  the setup described in Fig. \ref{fig:setup} by removing the SHG modules (brown curves in Fig. \ref{fig:phasenoise}A and B). 
It appears to be the dominant noise contribution of signals \circone and \circtwo in the 2 Hz${} - {}$50 Hz range but is found to be well rejected by the noise-rejection bench. 
Since AOMs are often used as actuators in optical frequency-control loops, their phase noise is automatically corrected. However, in ultra-stable laser setups where they only provide a frequency offset, our measurement implies that they need to be corrected for.
 
The common-mode rejection setup allows to isolate the phase noise of the frequency-doubled beam that is not  present in the pump beam, and  sets an upper bound on the residual phase noise added by the SHG process. The rejection setup shows a 30 dB noise reduction at 1 Hz, yielding a value ${\rm S}_\varphi(1\,{\rm Hz})=-38 \,{\rm dBrad^2 /Hz}$. A maximum rejection of 50 dB is reached around $f=50~\rm{mHz}$. The resulting integrated phase noise between 10~mHz and 30~Hz is compatible with interferometric gravitational wave detectors requirements \cite{Yeaton_Massey_2012}.

It is noteworthy that our results are qualitatively very similar to those shown in Ref. \cite{Yeaton_Massey_2012}, which are based on a post-treatment substraction of the pump laser noise in PPKTP crystals. When plot as phase noise, both spectra show a low-frequency floor followed by a steep slope (between $f^{-3}$ and $f^{-4}$), probably coming from low-frequency technical noise. The noise in Ref. \cite{Yeaton_Massey_2012} then shows the expected $f^{-1}$ slope, while our spectrum directly reaches a noise floor at frequencies above 800 Hz stemming from the low optical power. The intrinsic $f^{-1}$ noise of our crystals might be well below this noise floor.


Consistent with the overlapping of the ${\rm S}_\varphi(f)$ curves \circone and \circtwo in Fig. \ref{fig:phasenoise}A, the fractional frequency instabilities of \circone and \circtwo almost perfectly match. They yield fractional frequency instabilities below $10^{-15}$ at all averaging times, with the expected $\tau^{-1}$ slope corresponding to white and flicker phase noise. The total noise added when frequency doubling, including optical path length fluctuations and AOM and SHG modules temperature fluctuations, is low enough for optical ion clock setups. In particular, it is important to note that in our experiment, the free-space part of the interferometer was not shielded thermally nor acoustically; only the fiber splitter and the optical fiber inputs to the SHG modules were placed in an enclosure made of thick foam, providing basic thermal insulation. 

With the common-mode noise rejection setup, the fractional frequency instability drops to a value close to $ 10 ^{-17}$ at 1~s (see Fig. \ref{fig:phasenoise}B). This is better than what has ever been achieved so far in pulsed optical lattice clocks~\cite{Schioppo_2016} (purple dashed line in Fig. \ref{fig:phasenoise}B), and ultra-stable laser fractional frequency instabilities. It is still a value challenging to attain experimentally, as it implies  stabilization of the overall optical path to this level. This can be done by combining existing passive solutions (thermal and  acoustic shielding) and active solutions (fiber-stabilization and free-space path length stabilization)~\cite{Yeaton_Massey_2012}.
 
\section{Influence of external parameters}
 We have evaluated the noise induced by fluctuations of external parameters. 
 In particular, we measured the influence of pump laser frequency and intensity fluctuations and estimated the sensitivity of the module itself to variations of its environment.
 To do so, we force frequency, intensity, or temperature modulation at a frequency $f_0$ and we measure the power of the corresponding signal in the phase noise PSD ${\rm S}_\varphi(f_0)$. The values of the different conversion coefficients are summed up in Table \ref{tbl:XMPM}. 

  \begin{table}[b]
 	\caption{Conversion Coefficients of Various Physical Parameter Modulations into Phase Modulation.}
 	\label{tbl:XMPM}
 	\centering
 \begin{tabular}{|c|c|}
 	\hline 
Physical parameter &Conversion coefficient \\
 	\hline 
Pump laser frequency 	& $\alpha_{\rm FM-PM}=4(1)\,{\rm mrad/MHz}$  \\ 
Pump laser intensity 	& $\alpha_{\rm AM-PM}=0.4(1)\,{\rm mrad/\%}$ \\ 
 	\hline 
Module input intensity 	& $\beta_{\rm AM-PM}=26(5)\,{\rm mrad/\%}$ \\ 
Module temperature 	& $\beta_{\rm TM-PM}\approx 20(10)\,{\rm rad/K}$ \\ 
 	\hline 
 \end{tabular} 
\end{table}

 In the previous paragraphs, we assumed the frequency noise from the pump laser to be rejected by the interferometer. This  is rigorously true only if there is no delay between the two interferometer arms. In the case of a non-zero delay, the frequency noise of the pump laser is converted to phase noise on the interference output signal~\cite{Rubiola2005}. The frequency modulation to phase modulation (FM-PM) conversion for our setup was measured in order to know whether the influence  of the pump laser frequency noise could be neglected.
 
 A modulation of the ECDL frequency is performed by modulating the voltage of its piezoelectric transducer (PZT), at modulation frequencies from 10 to 400 Hz and for modulation depths from 15 to 70~MHz.
We found that the FM-PM conversion coefficient does not change significantly within this range of measurements, yielding a coefficient ${\alpha_{\rm FM-PM}=4(1)\,{\rm mrad/MHz}}$, consistent with the $\approx$ 20 cm delay between the two paths.
The influence of FM-PM conversion on the phase-noise spectra of Fig. \ref{fig:phasenoise}A would be visible  only for lasers with frequency stability $\sigma_y\geq 3\times 10^{-7}$ at 1~s, while the frequency stability of the free-running ECDL is expected to be typically $10^{-9}$ at 1~s. FM to PM conversion is thus negligible in the phase noise and stability measurements described above.
   
To evaluate the influence of pump laser intensity fluctuations, the laser intensity is modulated thanks to a liquid crystal modulator and the subsequent amplitude modulation to phase modulation (AM-PM) conversion is measured. 

It is modulated at the closest output of the ECDL, 
 with a modulation frequency of 10 -- 100~Hz and a depth between 1.5 and 6 \%.
 We measure a conversion coefficient for this common-mode amplitude modulation of ${\alpha_{\rm AM-PM}=0.4(1)\,{\rm mrad/\%}}$.  With the residual intensity noise (RIN) of the pump laser being lower than $-100\,{\rm dB/Hz}$ at all frequencies above 1~Hz, the contribution of the pump laser RIN to the measurements presented Fig \ref{fig:phasenoise} is estimated to be below $-130\,{\rm dBrad^2 /Hz }$, and can therefore be neglected. Using this conversion coefficient, we have also checked that the amplitude modulation induced by the displacement of the PZT in FM-PM conversion measurement is negligible. 

We have investigated the existence of AM-PM conversion in the SHG module itself. For this purpose, the laser intensity is modulated in one of the Mach-Zehnder arms, and the resulting phase modulation is measured between 1 and $10^4$~Hz and for modulation depths of $0.03 - 4\,\%$. We find a conversion coefficient ${\beta_{\rm AM-PM}=26(5)\,{\rm mrad/\%}}$. We have also measured the phase modulation induced by the liquid crystal modulator itself and found that its contribution represents at most 3~mrad/\%.
 
We also modulated the module temperature and measured a temperature modulation to phase modulation coefficient ${\beta_{\rm TM-PM}\approx 20(10)\,{\rm rad/K}}$. This is only a rough estimate, limited by long thermal time constants and consistent with independent evaluation from ${\rm LiNbO_3}$ properties.
 
The AM to PM conversion mechanism in the non-linear PPLN crystal could come from a dependence of  the crystal temperature on the light intensity. The pump beam intensity modulation would induce a temperature  modulation, which in turn would modulate the crystal length and cause a phase modulation.

\section{Conclusion}
For the first time, we have set an upper bound on residual phase noise induced by second harmonic generation in PPLN crystals. A novel noise rejection method allows real-time, physical substraction of the pump laser noise, leading to an optical phase noise below $- 35~{\rm dBrad^2 /Hz}$ at 1~Hz. Active locking of the optical path would give an even more stringent upper bound on excess phase noise \cite{Yeaton_Massey_2012}. Nevertheless, our results already show that  owing to their high conversion efficiency and their fractional frequency instability below $10^{-15}\tau^{-1/2}$, these PPLN modules are suitable for direct use in single-ion optical clocks. For more demanding applications such as optical lattice clocks or gravitational waves detection, a passive shielding of free-space optical paths and an active compensation of fiber-links is required to reach levels below $10^{-16}$.

\section*{Acknowledgments}

This work was supported by the Centre National d'\'Etudes Spatiales (R-S14/SU-0001-042), Agence Nationale de la Recherche (ANR-14-CE26-0031, ANR-10-LABX-48601, OSCILLATOR-IMP), and Région Bourgogne-Franche-Comté.

The authors would like to thank Ahmed Bakir and Baptiste Mar\'echal for technical support and Thomas Lévèque, François-Xavier Esnault and Vincent Giordano for fruitful discussions and comments.

\ifCLASSOPTIONcaptionsoff
  \newpage
\fi



\bibliographystyle{IEEEtran}
\bibliography{bibliographie2}




\end{document}